\documentstyle[12pt,epsf,epsfig]{article} 
\normalbaselines
\begin{document}
\bibliographystyle{unsrt}

\vbox {\vspace{6mm}} 
\def\bEQ{\vspace*{-1.9ex} \begin{equation}}  
\def\eEQ{\vspace*{-1.9ex} \end{equation}}    
\def\bEQA{\vspace*{-1.8ex} \begin{eqnarray}}  
\def\eEQA{\end{eqnarray} \\[-3.25ex]}  
\begin{center}
{\large \bf  
Photon adding and subtracting 
and Schr\"{o}dinger-cat generation \\[2mm]
in conditional output measurement on a beam splitter}\\[7mm]
D.--G. Welsch, M. Dakna, L. Kn\"{o}ll, and T. Opatrn\'{y}\\
{\it Theoretisch-Physikalisches Institut, 
Friedrich-Schiller-Universit\"{a}t Jena 
\\D-07743 Jena, Germany }
\end{center}

\vspace{2mm}

\begin{abstract}
The problem of photon adding and subtracting is studied,
using conditional output measurement on a beam splitter.
It is shown that for various classes of states the corresponding 
photon-added and -subtracted states can be prepared. Analytical 
results are presented, with special emphasis on photon-added and 
-subtracted squeezed vacuum states, which are found to represent 
two different types of Schr\"{o}dinger-cat-like states. Effects 
of realistic photocounting and Fock-state preparation are discussed.
\end{abstract}

\section{Introduction}
\label{sec1}
Quantum state engineering has been a subject of increasing interest
and various methods for generating highly nonclassical states of
optical fields have been proposed. A promising method for generating
novel quantum states has been conditional measurement. In fact,
when a system, such as a correlated two-mode optical field or a
correlated atom-field system, is prepared in an entangled state of
two subsystems and a measurement is performed on one subsystem, then
the quantum state of the other subsystem can be reduced to a new state. 
In particular, it turned out that conditional measurement on a 
beam splitter may be advantageously used for generating
new classes of quantum states \cite{Ban1}. 

Very interesting states are the so-called photon-added and 
photon-subtracted states that are obtained by repeated application 
of the photon creation and destruction operator, respectively, on a 
given state \cite{Ban1,Agarwal1,Dodonov1,Jones1}. In this paper we show 
that for various classes of states the corresponding photon-added 
and -subtracted states can be produced by means of conditional
measurement on a beam splitter. In particular, when a 
signal mode is mixed with a reference mode 
prepared in a photon-number state and in one of the output channels 
no photons are detected, then the mode in the other
output channel is prepared in a photon-added state. Accordingly,
a photon-subtracted state is produced when the reference mode is
in the vacuum and the conditional output measurement yields
a nonvanishing number of photons. 
Typical examples of states the method applies to
are thermal states, coherent states, squeezed states, 
displaced photon-number states and others. It is worth noting
that photon-added and photon-subtracted squeezed vacuum states 
represent two different classes of Schr\"{o}\-din\-ger-cat-like states. 
We analyze the states in terms of quadrature-component 
distributions and phase-space functions and calculate the probability
of producing them. Finally we discuss modifications that may be observed
under realistic experimental conditions.  
 
\section{Generation of photon-added and -subtracted states}
\label{sec2}
It is well known that the input--output relations at a lossless beam 
splitter can be treated within the SU(2) algebra. In the Schr\"{o}dinger 
picture, the output-density operator $\hat{\varrho}_{\rm out}$ can be 
obtained from the input--density operator $\hat{\varrho}_{\rm in}$ as
$\hat{\varrho}_{\rm out}$ $\!=$ $\!\hat{V}^{\dagger} \hat \varrho_{\rm in}  
\hat{V}$, where the operator $\hat{V}$ is given by $\hat{V}$ $\! =$
$e^{-i(\varphi_{T}-\varphi_{R}) \hat{L}_{3}}$ $\!e^{-2i\theta \hat{L}_{2}}$
$\!e^{-i(\varphi_{T}+\varphi_{R}) \hat{L}_{3}}$, with $\hat{L}_{2}$ $\!=$ 
$\!\textstyle\frac{1}{2i}(\hat{a}_{1}^\dagger\hat{a}_{2}$ $\!-$ 
$\!\hat{a}_{2}^\dagger\hat{a}_{1})$ and $\hat{L}_{3}$ $\!=$ 
$\! \textstyle\frac{1}{2}(\hat{a}_{1}^\dagger\hat{a}_{1}$ $\!-$ 
$\!\hat{a}_{2}^\dagger\hat{a}_{2})$. Let us assume that the input-state 
density operator can be written as
\bEQ
\hat \varrho_{\rm in}(n_{0}) = \hat \varrho_{{\rm in}1}
\otimes |n_0\rangle \langle\,n_0|,\quad n_0=0,1,2 ...
\label{1.04}
\eEQ
($\hat \varrho_{{\rm in}1}$, density operator of the mode in the first
input channel; $|n_{0}\rangle$, Fock state of the second input mode), 
then the output-state density operator $\hat{\varrho}_{\rm out}$ 
$\!\equiv$ $\!\hat{\varrho}_{\rm out}(n_{0})$ can be given by
\bEQA
\lefteqn{
\hat \varrho _{\rm out}(n_0) = \frac{1}{|T|^{2n_0}}
\sum_{n_{2}=0}^{\infty}\sum_{m_{2}=0}^{\infty}\sum_{k=0}^{n_0}
\sum_{j=0}^{n_0} (R^*)^{m_2+j} R^{n_2+k} 
}
\nonumber \\  &&  \times 
\frac{(-1)^{n_2+m_2} }{ \sqrt{k!j!m_2!n_2!} }
\sqrt{\! {n_0\choose k}\!{n_0\choose j}\!
{n_0\!-\!k\!+\!m_2\choose m_2}\!{n_0\!-\!j\!+\!n_2\choose n_2} }  
\nonumber \\  &&  \times \,
T^{\hat{n}_{1}}
{\hat{a}_{1}}^{m_2}({\hat{a}_{1}^\dagger})^{k}
\hat{\varrho}_{{\rm in}1}\hat{a}_{1}^{j}
({\hat{a}_{1}^\dagger})^{n_{2}}(T^*)^{\hat{n}_{1}}
\otimes 
|n_0-k+m_{2}\rangle\langle n_0-j+n_{2}|,
\label{1.0}
\eEQA
where $T$ $\!=$ $\! \cos\theta\;e^{i\varphi_{T}} $ and $R$ $\! =$ 
$\! \sin\theta\;e^{i\varphi_{R}} $, respectively, are the reflectance
and transmittance of the beam splitter. From Eq.~(\ref{1.0}) we see that 
the output modes are, in general, highly correlated. When the photon number 
of the mode in the second output channel is measured and $m_2$ photons are 
detected, then the mode in the first output channel is prepared in a quantum 
state whose density operator $\hat{\varrho}_{{\rm out}1}(n_{0},m_{2})$ 
reads as
\bEQA
\hat{\varrho}_{{\rm out}1}(n_0,m_2)
= \frac{\langle m_{2} |\hat{\varrho}_{\rm out}(n_0) | m_{2} \rangle}
{{\rm Tr}_{1}( \langle m_{2} |
\hat{\varrho}_{\rm out}(n_{0}) | m_{2} \rangle)} \,.
\label{1.1}
\eEQA
The probability of such an event is given by
\bEQA
\lefteqn{
\hspace*{-4ex}
P(n_0,m_2)=
{\rm Tr}_{1}( \langle m_{2} |
\hat{\varrho}_{\rm out}{(n_0)} | m_{2} \rangle )
= \!\! \sum_{n_{1}=\mu-\nu}^{\infty}
\sum_{j=\mu}^{n_0}\sum_{k=\mu}^{n_0}\;
\!\!|R|^{2(j+k-\nu)}|T|^{2(n_1+\nu-n_0)}
}
\nonumber \\ && \hspace{-4ex}\times \,
\frac{(-1)^{j+k}n_0!\,n{_1}!}{(n_1+\nu)!(n_0-\nu)!}
{n_0-\nu\choose j-\nu} {n_0-\nu\choose k-\nu}
{n_1+j\choose j}{n_1+k\choose k}
\langle n_{1} | \hat{\varrho}_{{\rm in}1} | n_{1} \rangle,
\label{1.6}
\eEQA
where $\nu=n_0-m_2$, $\mu=\max(0,\nu)$.

   From Eqs.~(\ref{1.0}) and (\ref{1.1}) we find that when a signal mode 
prepared in a state $\hat{\varrho}_{{\rm in}1}$ $\!=$ 
$\!\sum_{\Phi} \tilde p_{\Phi} \, | \Phi\rangle\langle \Phi  |$ 
($\sum_{\Phi}\tilde p_{\Phi}$ $\!=$ $\!1$, $0$ $\!\leq$ $\!\tilde p_{\Phi}$
$\!\leq$ $\!1$) is mixed with a mode prepared in a Fock state and a 
zero-photon conditional measurement is performed in the second output
channel ($n_{0}$ $\!>$ $\!0$, $m_{2}$ $\!=$ $\!0$), then the mode in the 
first output channel is prepared in the state 
\bEQ
\hat{\varrho}_{{\rm out}1}(n_0\neq 0,m_{2}=0)
= \sum_{\Phi} \tilde p_{\Phi} | \Psi_{n_0,0} \rangle 
\langle \Psi_ {n_0,0} |, \quad
|\Psi_{n_0,0}\rangle=
{\cal N}_{n_0,0}^{-1/2}
\, (\hat{a}^\dagger_1)^{n_0}\,T^{\hat{n}_1} |\Phi\rangle .
\label{1.4ma}
\eEQ
As can be seen, the state $|\Psi_{n_0,0} \rangle$ is a photon-added
state, $n_{0}$ photons beeing added to the state $T^{\hat{n}_1} 
|\Phi\rangle$. For a number of classes of states (e.g., thermal, 
coherent, squeezed, displaced Fock states etc.) the states 
$|\Phi\rangle$ and $T^{\hat{n}_1} |\Phi\rangle$ belong to 
the same class.In this case conditional output measurement provides 
us with a method for generating photon-added states of the class of 
states to which a signal-mode quantum state $|\Phi\rangle$ belongs.
Similarly, when the second input mode is in the vacuum state and a 
nonvanishing number of photons is detected ($n_{0}$ $\!=$ $\!0$, 
$m_{2}$ $\!>$ $\!0$), then a photon-subtracted state is produced:
\bEQ
\hat{\varrho}_{{\rm out}1}(n_0=0,m_{2}\neq 0)
= \sum_{\Phi} \tilde p_{\Phi}| \, \Psi_{0,m_2} \rangle \big\langle 
\Psi_ {0,m_2}| , \quad
|\Psi_{0,m_2}\rangle=
{\cal N}_{0,m_2}^{-1/2}
\, (\hat{a}_1)^{m_{2}}\,T^{\hat{n}_1} |\Phi\rangle .
\label{1.4m}
\eEQ
In Eqs.~(\ref{1.4ma}) and (\ref{1.4m}), ${\cal N}_{0,m_2}$ and 
${\cal N}_{n_0,0}$ are normalization constants.
The probabilities of observing the photon-added and photon-subtracted
states, respectively, are found from Eq.~(\ref{1.6}) to be
\bEQ
P(n_{0}) \equiv P(n_0\neq 0 ,m_2=0)
=|R|^{2n_0}\sum_{n_1=0}^{\infty}
|T|^{2n_1}
{n_1+n_0\choose n_0 }
\langle n_{1} | \hat{\varrho}_{{\rm in}1}| n_{1} \rangle 
\label{M1a}
\eEQ
and
\bEQ
P(m_2)\equiv P(n_0=0,m_2\neq 0)=|R|^{2m_{2}}
\!\! \sum_{n_{1}=0}^{\infty}|T|^{2n_{1}}
{n_{1}+m_2\choose m_{2}}  \!
\langle n_{1}+m_2 | \hat{\varrho}_{{\rm in}1} | n_{1}+m_2 \rangle .
\label{M1}
\eEQ
In order to illustrate the method, let us restrict attention to 
photon-added and photon-subtracted squeezed-vacuum states.

\section{Photon-added and -subtracted squeezed vacuum states}
\label{sec3}
When the input state $|\Phi\rangle$ is a squeezed vacuum state,
\bEQA
|\Phi\rangle =
|0\rangle_{\rm s} = (1-|\kappa|^{2})^{1/4}\sum_{n=0}^{\infty}
\frac{[(2n)!]^{1/2}}{2^{n}\,n!}\,
\kappa^{n}|2n\rangle,
\label{1.2}
\eEQA
then the conditional output states $|\Psi_{n_{0},0}\rangle$, Eq.~(\ref{1.4ma}),
and $|\Psi_{0,m}\rangle$, Eq.~(\ref{1.4m}), are given by
\bEQA
|\Psi_{n_0,0}\rangle =
{\cal N}_{n_0,0}^{-1/2}
\sum_{n=n_0}^{\infty} c_{n,n_0,0} |n\rangle,
\quad c_{n,n_0,0}=\frac{\sqrt{n!}}
{\Gamma\!\left[\frac{1}{2}(n-n_0)+1\right]}
\,{\textstyle\frac{1}{2}\left[1+(-1)^{n-n_0}\right]
(\frac{1}{2} \kappa' )^{\frac{n-n_0}{2}}} ,
\label{M7a}
\eEQA
and
\bEQA
|\Psi_{0,m}\rangle = 
{\cal N}_{0,m}^{-1/2}
\sum_{n=0}^{\infty}c_{n,0,m}  |n\rangle,\quad c_{n,0,m}=
\frac{(m+n)!}
{\Gamma\!\left[\frac{1}{2}(n+m)+1\right]\sqrt{n!}}
\,{\textstyle\frac{1}{2}\left[1+(-1)^{n+m}\right]
(\frac{1}{2}\kappa')^{\frac{n+m}{2}}} ,
\label{M7}
\eEQA
where $\kappa'$ $\!=$ $\!T^2\kappa$, and the normalization constants 
are derived to be 
\bEQA
{\cal N}_{0,m}
=\frac{{|\kappa'|}^{2m}}{(1-{|\kappa'|}^2)^{m+1/2}}
\sum_{k=0}^{\left[\frac{m}{2}\right] }
\frac{\left(m!\right)^2(2|\kappa'|)^{-2k}}
{(m-2k)! \left( k! \right)^{2}},\quad
{\cal N}_{n_0,0}
= n_0!\,{\rm F}\!\left[\textstyle\frac{1}{2}(n_0+1),
\textstyle\frac{1}{2}(n_0+2),1;|\kappa'|^2\right].
\eEQA
For notational convenience we omit the subscripts $1$ and $2$ introduced 
above to distinguish between the two output channels.
The probabilities of producing the states $|\Psi_{n_0,0}\rangle$ and 
$|\Psi_{0,m}\rangle$ are
\vspace{0.5ex}
\bEQ
P(n_0) = |R|^{2n_0} \sqrt{1-|\kappa|^2}
\, {\rm F}\!\left(n_0+1,\textstyle\frac{1}{2},1;|\kappa'|^2\right)
\label{M13a}
\eEQ
\vspace{0.5ex}
and
\vspace{0.5ex}
\bEQ
P(m)=
\frac{2^{-m} |R|^{2m}\sqrt{1-|\kappa|^2}}
{(1-{|\kappa'|}^2)^{m+1/2}|T|^{2m}}
\sum_{k=0}^{\left[ \frac{m}{2} \right]}
\frac{m!(2|\kappa'|)^{m-2k}}{(m-2k)!(k!)^{2}},
\\[1ex]
\label{M13}
\eEQ
respectively [F$(a,b,c;z)$, hypergeometric function ].

The states $|\Psi_{n_0,0}\rangle$ and $|\Psi_{0,m}\rangle$ are
Schr\"{o}dinger-cat-like states \cite{Dakna1}, i.e., superpositions 
of two (macroscopically distinguishable) ``component'' states 
that are well localized in the phase space. The properties of 
$|\Psi_{n_0,0}\rangle$ and $|\Psi_{0,m}\rangle$ can be seen from the
quadrature-component distributions $p_{n_0}(x,\varphi|0)$ and 
$p_0(x,\varphi|m)$:
\bEQA
p_{n_0}(x,\varphi|0)&=&
\frac{1}{{\cal N}_{n_0,0} \sqrt{\pi \Delta^{n_0+1}}2^{n_0}}
\,\exp\!\left( - \frac{1-|\kappa'|^{2}}{\Delta} \, x^{2} \right)
\left |
{\rm H}_{n_0}\!\left[
\sqrt{(1+{\kappa'}^\ast e^{i2\varphi})/\Delta}\,x
\right]\right |^2
,\\
p_0(x,\varphi|m)&=&
\frac{|\kappa' |^{m}}{{\cal N}_{0,m} \sqrt{\pi \Delta^{m+1}} \, 2^{m}}
\exp\!\left( - \frac{1-|\kappa'| ^2}{\Delta} \, x^{2} \right)
\left |
{\rm H}_m\!\left[
\sqrt{(-{\kappa'}^\ast e^{i2\varphi}-|\kappa'|^2)/\Delta}\,x
\right]\right |^2 ,
\label{F4}
\eEQA
where $\varphi_{\kappa'}\!=\!\varphi_{\kappa}\!+\!2\varphi_T$, and
$ \Delta$ $\!=$ $\!1$ $\!+$ $\!|\kappa'|^2$ $\! +$ 
$\!2 |\kappa'|\cos(2\varphi-\varphi_{\kappa'})$ [${\rm H}_n(z)$, Hermite
polynomial]. The corresponding
Wigner functions are calculated to be
\bEQA
\label{W4a}
\lefteqn{
W_{n_0}(x,p|0)=
\frac{|\kappa'|^{n_0}}{\pi{\cal N}_{n_0,0}2^{n_0}|1+\kappa'|^{2n_0+1}}
\left(\frac{2}{\lambda+\lambda^\ast}\right)^{n_0+1/2}
\exp\!\left(-\frac{2|\lambda|^2}{\lambda+\lambda^\ast}
\left|x+i\frac{p}{\lambda}\right|^2\right)
}
\nonumber \\ && \hspace{15ex} \times \,
\sum_{k=0}^{n_0}{n_0\choose k}^2 \!k!\left(\frac{-2}{|\kappa' |}\right)^k
\left |{\rm H}_{n_0-k}\!\left[i\sqrt{\frac{2\lambda^2(1+\lambda^\ast)}
{(1-\lambda)(\lambda+\lambda^\ast)}}
\left(x+i\frac{p}{\lambda}\right)\right]\right |^2\!\!,
\eEQA
\bEQA
\label{W4}
\lefteqn{W_0(x,p|m)=
\frac{|\kappa'|^m}{\pi{\cal N}_{0,m}2^m|1+\kappa'|^{2m+1}}
\left(\frac{2}{\lambda+\lambda^\ast}\right)^{m+1/2}
\exp\!\left(-\frac{2|\lambda|^2}{\lambda+\lambda^\ast}
\left|x+i\frac{p}{\lambda}\right|^2\right)
}
\nonumber \\ && \hspace{15ex} \times \,
\sum_{k=0}^{m}{m\choose k}^2 \!k!\left(-2|\kappa' |\right
)^k\left |{\rm H}_{m-k}\!\left[i\sqrt{\frac{2\lambda^2(1-\lambda^\ast)}
{(1+\lambda)(\lambda+\lambda^\ast)}}
\left(x+i\frac{p}{\lambda}\right)\right]\right |^2\!\!,
\eEQA
where $\lambda$ $\!=$ $\!(1$ $\!-$ $\!\kappa')$ $\!/$
$\!(1$ $\!+$ $\!\kappa')$, and the Husimi functions read as
[$\alpha$ $\!=$ $\!2^{1/2}(x$ $\!+$ $\!ip)$]
\vspace{1ex}
\bEQ
\label{Q3a}
Q_{n_0}(x,p|0)=\frac{|\alpha|^{2n_0}e^{-|\alpha|^2}}{2\pi{\cal N}_{n_0,0}}
\exp\!\left[{\textstyle\frac{1}{2}}
({\kappa'}^\ast\alpha^2+\kappa'{\alpha^\ast}^2) \right ],
\eEQ
\vspace{0.5ex}
\bEQ
\label{Q3}
Q_{0}(x,p|m)=\frac{|\kappa'|^{m}e^{-|\alpha|^2}}{2\pi{\cal N}_{0,m}2^m}
\exp\!\left[{\textstyle\frac{1}{2}}
({\kappa'}^\ast\alpha^2+\kappa'{\alpha^\ast}^2) \right ]
\left |
{\rm H}_m\left(\sqrt{-\textstyle\frac{1}{2}{\kappa'}^\ast}\alpha\right)
\right|^2.
\eEQ

\section{State mixing in real experiments}
\label{sec4}
Let us first address the problem of realistic photon-number measurement 
in photon-subtracted state generation. Since highly efficient
and precisely discriminating photodetectors are not available at present,
multichannel photon chopping \cite{Paul1} may be used. For a $2N$-port apparatus 
the probability of recording $k$ coincident events when $m$ photons are
present is given by 
\vspace{0.2ex}
\bEQ
\label{T2}
\tilde P_{N,\eta}(k|m) = \sum_{l} \tilde P_{N}(k|l) \, M_{l,m}(\eta) 
\eEQ 
($\eta$, detection efficiency), where $\tilde P_{N}(k|m)$ 
$\!=$ $\!{M}_{l,m}(\eta)$ $\!=$ $\!0$ for $k,l$ $\!>$ $\!m$, and 
\bEQ
\label{T1}
\tilde P_{N}(k|m) = \frac{1}{N^{m}} {N \choose k}
\sum_{l=0}^{k}(-1)^{l} {k \choose l} (k - l)^{m}
\quad {\rm for} \quad k\le m,  
\label{4.1}
\\[1.5ex]
\eEQ 
\bEQ
{M}_{l,m}(\eta) = {m \choose l} \eta^{l} (1 - \eta )^{m-l}
\quad {\rm for} \quad l\le m.
\\[1ex]
\label{4.bb}
\eEQ 
Since detection of $k$ coincident events can result from various numbers
of photons, $m$, the conditional output state is in general a statistical 
mixture. In place of $|\Psi_{0,m}\rangle$, Eq.~(\ref{1.4m}), we have
\bEQ
\label{T3}
\hat \varrho
= \sum_{m}  P_{N,\eta}(m|k) \,|\Psi_{0,m} \rangle\langle \Psi_{0,m} |,
\\[1ex]
\eEQ 
where the conditional probability $P_{N,\eta}(m|k)$ can be obtained as,
on using the Bayes rule, 
\noindent
\bEQ
\label{T4}
P_{N,\eta}(m|k) = \frac{1}{\tilde P_{N,\eta}(k)} \tilde P_{N,\eta}(k|m)
P(m).
\\[.5ex]
\eEQ 
Here $P(m)$ is the prior probability (\ref{M13}) of $m$ photons 
being present, and
\vspace*{.5ex}
\noindent
\bEQ
\tilde P_{N,\eta}(k) = \sum_{m} \tilde P_{N,\eta}(k|m) P(m) 
\eEQ 
is the prior probability of recording $k$ coincident events. In Figs.~1(a) 
and 1(b), examples of the resulting quadrature distribution $p_0(x,\varphi|k)$ 
$\!=$ $\!\sum_{m}$ $\!P_{N,\eta}(m|k)$ $\!p_{0}(x,\varphi|m)$ and
the Wigner function $W_0(x,p|k)$ $\!=$ $\!\sum_{m}$ $\!P_{N,\eta}(m|k)$  
$\!W_{0}(x,p|m)$, respectively, are plotted.
\begin{figure}[tbh]
\hspace{-5ex}
\begin{minipage}[b]{0.38\linewidth}
\centering\epsfig{figure=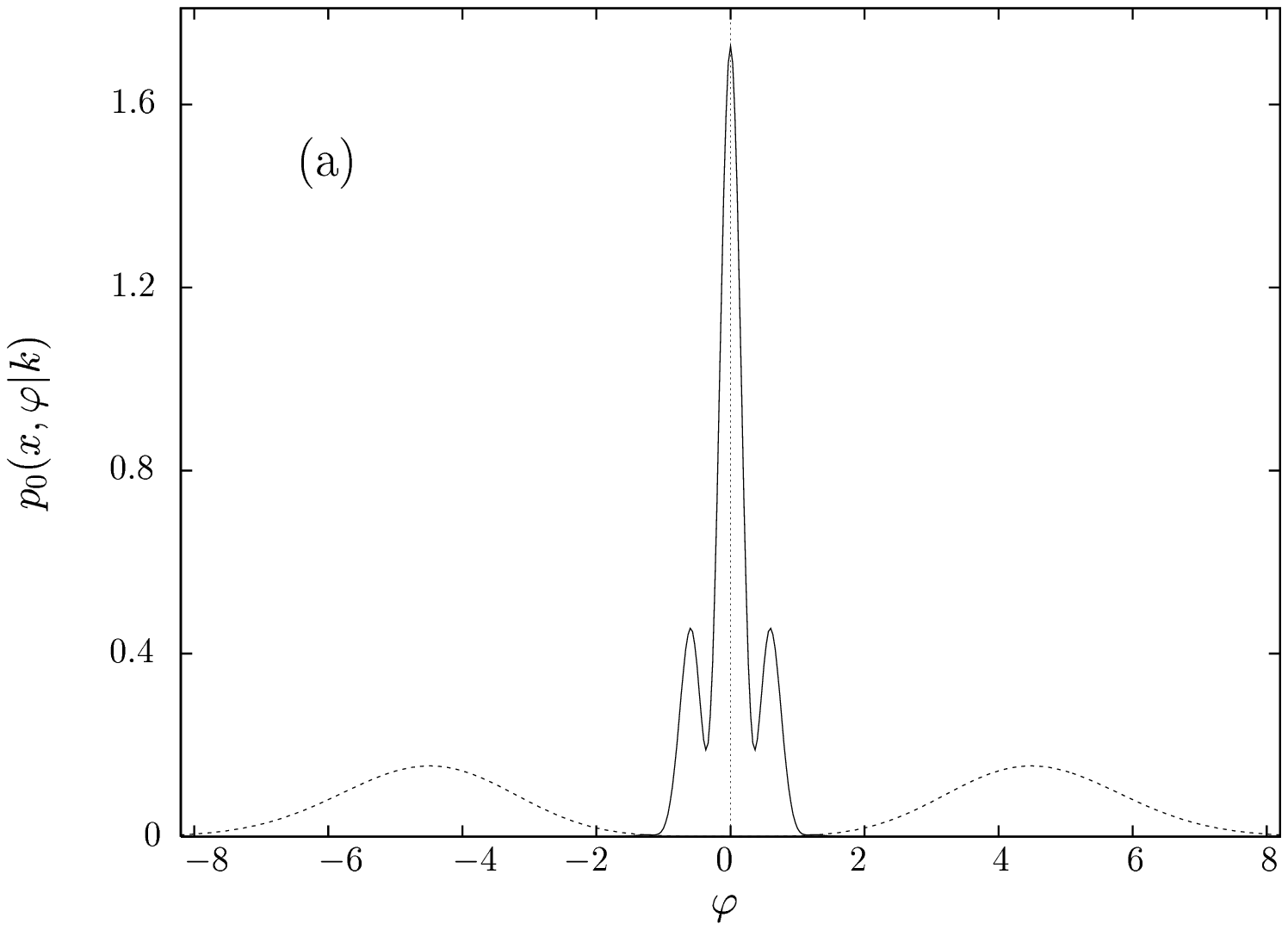,width=.9\linewidth}
\end{minipage} \hfill 
\begin{minipage}[b]{0.36\linewidth}
\centering\epsfig{figure=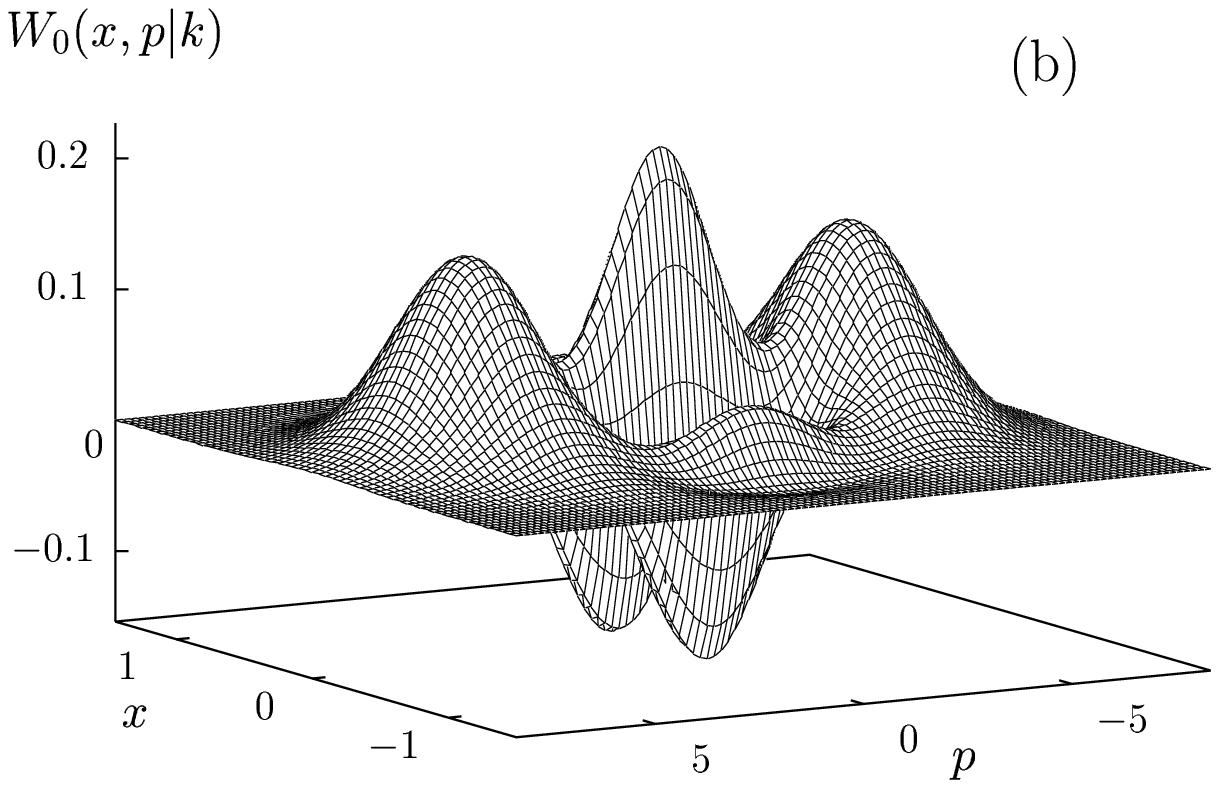,width=\linewidth}
\end{minipage} \hspace{1ex}
\begin{minipage}{0.27\linewidth}
\vspace*{-25ex}
\footnotesize{
FIG.~1.~Quadrature distribution (a) [for the phase parameters 
$\varphi$ $\! =$ $ \! 0$ (full line) and $\varphi$ $\! =$ $\! \pi /2$ 
(broken line)] and the Wigner function (b) of a photon-subtracted 
squeezed vacuum with realistic photodetection 
($N$ $\!=$ $\!20$, $\eta$ $\! =$ $ \!0.95$)
for $k$ $\! =$ $ \! 4$ coincident events, and $|\kappa|$ $ \! =$ $ \! 0.77$, 
$|T|^2$ $ \! =$ $ \! 0.9$ ($\kappa'$ $ \! =$ $ \! -0.7$). 
$P_{N,\eta}(k)$ $\! =$ $ \!0.07\%$}
\end{minipage}
\end{figure} 

In order to produce photon-added states, the second input mode must be 
prepared in a Fock state. This is a nontrivial experimental task, and a 
number of proposals have been made. Therefore it may be more realistic 
to consider a sub-Poissonian statistical mixture of Fock states
rather than a pure state. Let us 
return to Eq.~(\ref{1.04}) and suppose that $\hat \varrho_{\rm in}$ 
$\!=$ $\!\hat \varrho_{{\rm in}1}$ $\!\otimes$ $\!\hat \varrho_{{\rm in}2}$,
with $\hat \varrho_{{\rm in}2}$ $\!=$ $\!\sum_{n_{0}}$  
$\!\tilde p_{n_{0}}$ $\!|n_0\rangle\langle n_0|$. To give an
example, we assume that $\tilde p_{n_{0}}$ is a binomial probability 
distribution,
\bEQ
\tilde p_{n_{0}}
= {N\choose n_0} p^{n_0}(1-p)^{N-n_0} \quad {\rm if} \quad n_{0}\leq N, 
\\[1ex]
\label{AD1}
\eEQ
and $\tilde p_{n_{0}}$ $\!=$ $\!0$ elsewhere ($0$ $\!<$ $\!p$ $\!<$ $\!1$).
We then find that $|\Psi_{n_{0},0}\rangle$, Eq.~(\ref{1.4ma}), must be
replaced with the mixed state
\bEQ
\hat{\varrho}
= \sum_{n_{0}} \tilde p_{n_{0}} \, 
|\Psi_{n_0,0} \rangle \big\langle \Psi_ {n_0,0}| .
\label{AD5}
\eEQ
Accordingly, the probability of detecting the state is the average 
of $P(n_{0})$, Eq.~(\ref{M13a}), i.e., $P$ $\!=$ $\!\sum_{n_0}$
$\! \tilde p_{n_{0}} P(n_{0})$. Examples of the resulting 
quadrature-component distribution $p(x,\varphi|0)$ 
$\!=$ $\!\sum_{n_{0}}$ $\!\tilde p_{n_{0}}$ $\!p_{n_{0}}(x,\varphi|0)$
and the Wigner function $W(x,p|0)$ $\!=$ $\!\sum_{n_{0}}$ 
$\!\tilde p_{n_{0}}$ $\!W_{n_{0}}(x,p|0)$, respectively, 
are plotted in Fig.~2(a) and (b).
\vspace{1ex}
\begin{figure}[tbh]
\hspace{-5ex}
\begin{minipage}[b]{0.38\linewidth}
\centering\epsfig{figure=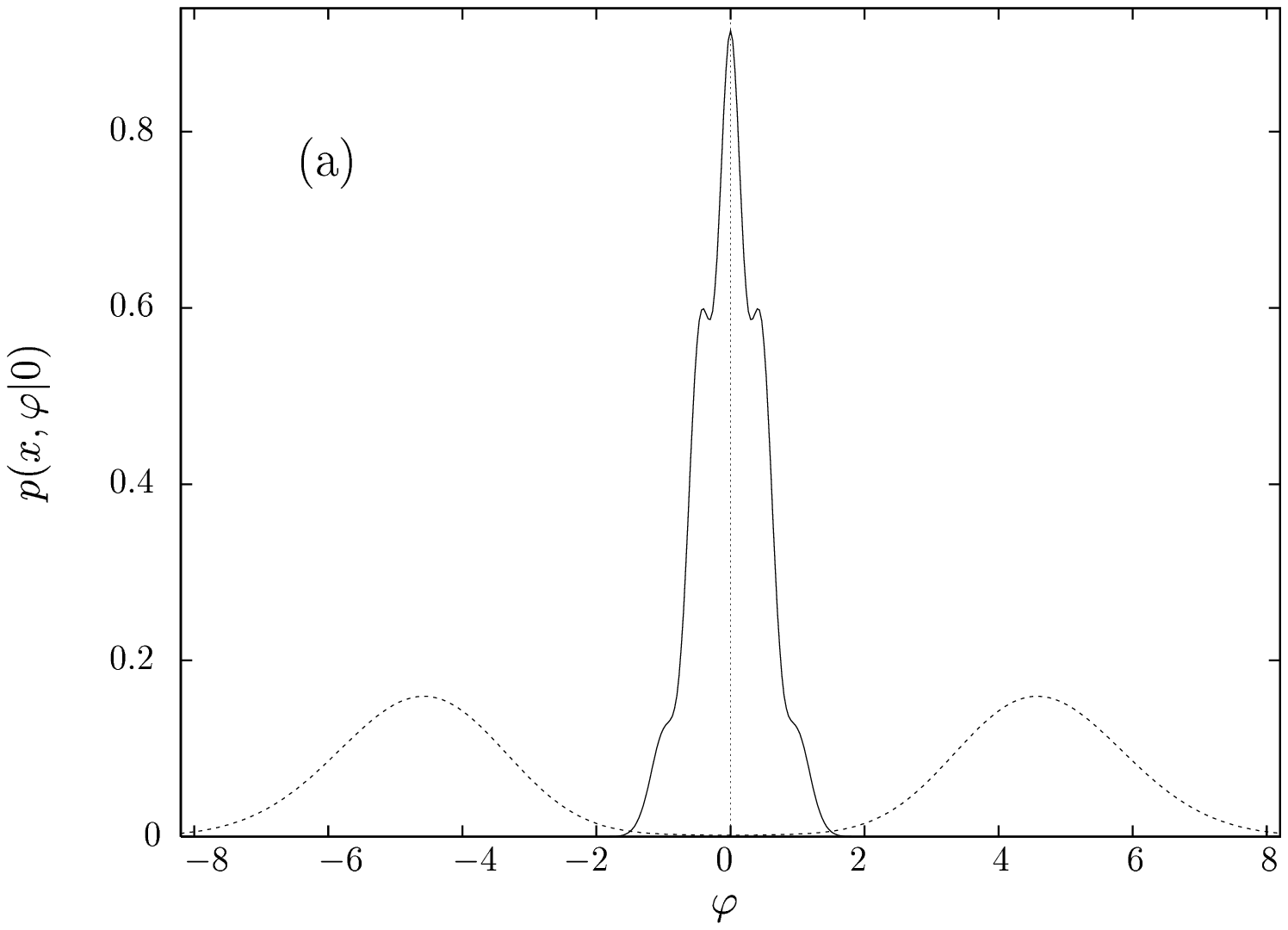,width=.9\linewidth}
\end{minipage} \hfill
\begin{minipage}[b]{0.36\linewidth}
\centering\epsfig{figure=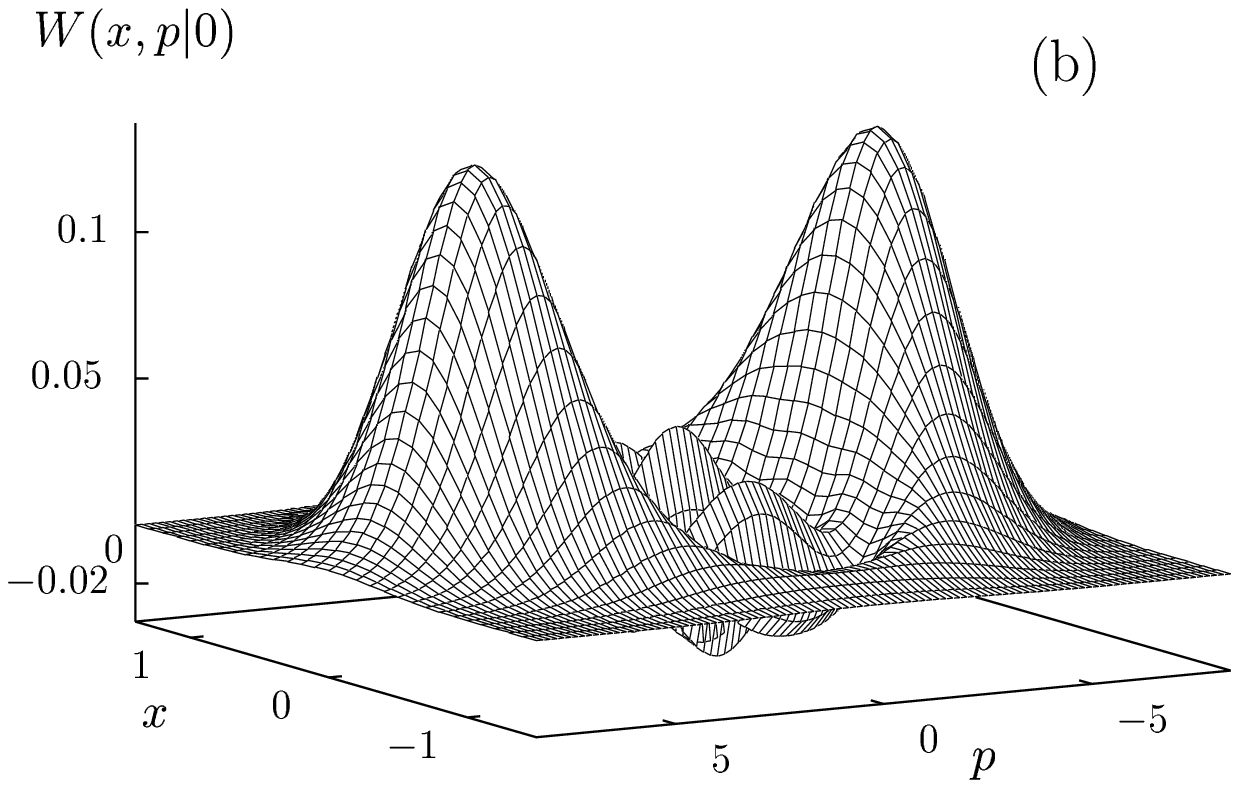,width=\linewidth}
\end{minipage} \hspace{1ex}
\begin{minipage}{0.26\linewidth}
\vspace{-23ex}
\footnotesize{FIG.~2.~Quadrature distribution (a) [for the phase parameters 
$\varphi$ $ \! =$ $\! 0$ (full line) and $\varphi$ $\! =$ $ \! \pi /2$ 
(broken line)] and the Wigner function (b) of a photon-added squeezed 
vacuum with realistic Fock-state preparation 
[$p$ $\! = $ $\! 0.8$ and $N$ $\! =$ $\! 4$ in 
Eq.~(\protect\ref{AD1})] for $|\kappa|$ $\!=$ $\!0.77$, $|T|^2$ $\! =$
$ \!0.9$ ($\kappa'$ $ \! =$ $\! -0.7$). 
$P$ $\! =$ $\! 0.02\%$}
\end{minipage}
\end{figure}

Figures.~1 and 2 reveal that -- apart from some smearing -- 
the non-classical Schr\"{o}dinger-cat-like features of 
photon-subtracted and photon-added squeezed vacuum states can still 
be preserved even under realistic experimental conditions. 
In particular it is seen that the typical quantum interferences 
can be observed.
    
\section{Summary and Conclusion}
\label{sec5}
We have studied the
feasibility of generating photon-added and photon-subtracted states
of a single-mode optical field via conditional output measurement on a 
beam splitter, with special emphasis on photon-added
and -subtracted squeezed vacuum states. When a squeezed vacuum and an
ordinary vacuum are mixed by a beam splitter and in one of the output 
channels a nonvanishing number of photons is detected, then a
photon-subtracted squeezed vacuum in the other output channel is
prepared, which shows all the features of 
a Schr\"{odinger}-cat-like state. Another class of Schr\"{odinger}-cat-like 
states can be prepared when photons are not subtracted from a squeezed vacuum 
but added to it, which is the case when a squeezed vacuum is mixed with a 
photon-number state and a zero-photon conditional output measurement is 
performed. We have derived the quadrature-component distributions and
the Wigner and Husimi functions for the two classes of states. Further
we have given the probabilities of preparing the states.

It is worth noting that the component states of a Schr\"{o}dinger-cat-like
state of the type of a photon-added squeezed vacuum can be regarded as
non-Gaussian squeezed coherent states that tend to the familiar
Gaussian squeezed coherent states (two-photon coherent states) as the 
number of added photons becomes sufficiently
large. In contrast to photon adding, the component states of a 
Schr\"{o}dinger-cat-like state of the type of a photon-subtracted 
squeezed vacuum are, in a very good approximation, Gaussian squeezed
coherent states which with increasing number of subtracted photons
approach ordinary coherent states.  

With regard to realistic experimental conditions, we have considered
multichannel photon-number detection in photon-subtracted state
generation as well as sub-Poissonian photon-number input statistics
in photon-added state generation. As expected, realistic photon-number
measurement smears the interference structure in the photon-subtracted
squeezed vacuum, and a similar effect is observed in the
photon-added squeezed vacuum when one allows for an input mode
prepared in a sub-Poissonian statistical mixture of photon-number
states in place of a pure Fock state. 
Nevertheless, the typical properties of Schr\"{o}dinger-cat-like states
can still be found even under realistic conditions.

\vspace{2ex}
\noindent {\bf Acknowledgment}
This work was supported by the Deutsche Forschungsgemeinschaft.

\vspace{15mm}

\end{document}